\documentclass[letterpaper, 10pt, twocolumn]{article}

\usepackage{graphicx} 
\usepackage{geometry}
\usepackage{authblk}

\usepackage{float}
\usepackage{amsmath}

\usepackage{soul}
\usepackage[dvipsnames]{xcolor}

\begin{document}

\author[1]{Grant M. Brodnik\thanks{grant.brodnik@nist.gov}}
\author[2]{Grisha Spektor}
\author[1,3]{Lindell M. Williams}
\author[1,3]{Jizhao Zang}
\author[1]{Alexa R. Carollo}
\author[1,3]{Atasi Dan\thanks{Present address: Micron Technology, Boise, ID}}
\author[1]{Jennifer A. Black\thanks{Present address: Vescent Technologies, Golden, CO}}
\author[2]{David R. Carlson}
\author[1,3]{Scott B. Papp}

\affil[1]{Time and Frequency Division, National Institute of Standards and Technology, Boulder, CO USA}
\affil[2]{Octave Photonics, Louisville, CO 80027, USA}
\affil[3]{Department of Physics, University of Colorado, Boulder, CO 80309, USA}

\title{Monolithic 3D integration of tantalum pentoxide photonics on arbitrary substrates}

\date{}
\maketitle

\noindent \textbf{The photonics landscape encompasses a wide scope of material platforms, each optimized for specific functionalities, yet no platform meets the demands of all current and evolving photonic applications. While combining integrated photonics materials enhances overall capability—such as unifying nonlinear optics, low-loss passive devices, and electro-optics—material and process compatibility remains a major challenge. We introduce full-wafer, monolithic 3D integration of tantalum pentoxide (Ta$_2$O$_5$, hereafter tantala)  photonics onto arbitrary substrates, which we explore here with thin-film lithium niobate (LN) on silicon. Tantala's unique properties, importantly room-temperature deposition, low-temperature annealing, and low stress in thick films optimized for phase matching, make it well suited for monolithic 3D integration without compromising substrate performance or compatibility. We demonstrate low-loss, high-quality-factor microresonators and nanophotonics in tantala, robust quasi-phase-matching in poled LN waveguides, and efficient 3D interlayer routing. This enables us to demonstrate a rich palette of nonlinear frequency conversion processes, including $\chi^{(3)}$ optical parametric oscillation (OPO) and soliton microcomb generation in tantala microresonators and photonic-crystal resonators, $\chi^{(2)}$ second-harmonic generation (SHG) in periodically poled LN, and combinations thereof. Monolithic 3D integration with tantala opens a new paradigm for scalable, multifunctional photonic systems, enabling visible, near-IR, and nonlinear operation into existing photonic infrastructure.}

\section{Introduction}

The advancement of integrated photonics has the potential to fuel innovation across a broad range of applications including metrology \cite{spencer_optical-frequency_2018}, telecommunications \cite{pfeifle_coherent_2014}, visible-light consumer electronics \cite{raval_integrated_2018}, and quantum technologies \cite{alexander_manufacturable_2025}. These examples highlight the diverse functional demands placed on photonic systems, requiring laser sources, frequency combs, high-speed modulators, and quantum light sources in single platforms. For higher performance and novel capabilities that meet these demands, recent efforts have focused on bringing nonlinear optics on-chip. 
Third-order nonlinear ($\chi^{(3)}$) platforms such as silicon nitride (SiN) \cite{pfeiffer_photonic_2016}, tantalum pentoxide (tantala) \cite{jung_tantala_2021}, and aluminum nitride \cite{jung_aluminum_2016}, support a range of efficient nonlinear phenomena including frequency-comb generation, supercontinuum generation, soliton dynamics, and broadband wavelength conversion.
In parallel, thin-film second-order nonlinear ($\chi^{(2)}$) systems using materials like lithium niobate (LN) \cite{desiatov_ultra-low-loss_2019,zhu_integrated_2021,boes_lithium_2023}, lithium tantalate \cite{wang_lithium_2024}, and barium titanate (BTO) \cite{karvounis_barium_2020} enable high-bandwidth electro-optic modulation \cite{wang_integrated_2018} and second harmonic generation (SHG), often through quasi-phase-matched poled waveguides \cite{wang_ultrahigh-efficiency_2018,jankowski_ultrabroadband_2020,younesi_periodic_2021,park_high-efficiency_2022}. 
Nonetheless, despite significant progress in both established and emerging platforms, it is widely recognized that no single material can satisfy all functional requirements. 
As a result, current efforts are increasingly focused on strategies that combine integrated-photonics platforms with high-performance nonlinear materials, in particular those supporting broadband low-loss across the visible and shortwave infrared (SWIR), to unlock the potential of integrated photonics.

Heterogeneous integration has emerged as one strategy for uniting complementary photonic materials.
By bonding entirely separate substrates with distinct materials for optical, electronic, and structural properties, such as integrating III–V semiconductors with low-loss passive and nonlinear waveguides, this approach enables the realization of complex, multifunctional photonic systems \cite{komljenovic_photonic_2018,fathpour_heterogeneous_2018,nader_heterogeneous_2025}.
Recent demonstrations have combined III–V gain media with nonlinear waveguides to realize integrated amplifiers and tunable lasers \cite{beeck_iiiv--lithium_2021,snigirev_ultrafast_2023} and microresonator frequency combs \cite{xiang_laser_2021} pumped by on-chip lasers.
Other efforts have leveraged techniques like micro-transfer printing \cite{vanackere_heterogeneous_2023} and direct bonding \cite{chang_heterogeneous_2017,ghosh_wafer-scale_2023,churaev_heterogeneously_2023} to incorporate thin-film lithium niobate onto SiN platforms. These developments are promising within select material ecosystems, though the need for recurring customization of fabrication to ensure material and process compatibility poses a significant barrier to scalability and therefore broader adoption across photonics.

Monolithic integration presents a streamlined and scalable path to combine materials and realize diverse functionalities within a unified platform and fabrication process. 
Originating in microelectronics, it involves sequential fabrication of multiple device layers on a common substrate within a cohesive process flow. 
This approach underpins complementary metal–oxide–semiconductor technology, where transistors, interconnects, and memory are vertically integrated at the wafer level to deliver dense and cost-effective circuits. 
In photonics, early monolithic efforts on indium phosphide (InP) platforms \cite{nagarajan_large-scale_2005} demonstrated integration of active and passive components like lasers, modulators, and photodetectors into a single chip, enabling compact optical transceivers and high-bandwidth communication systems. 
Silicon photonics subsequently advanced monolithic strategies by incorporating vertically stacked silicon-on-insulator (SOI) waveguide layers to increase routing density and device functionality within foundry-scale processes \cite{stojanovic_monolithic_2018}. 
More recently, monolithic three-dimensional photonic integration has emerged, particularly in SOI and silicon nitride (SiN) platforms, where multiple photonic layers are stacked and interconnected \cite{sacher_multilayer_2015,bose_anneal-free_2024}.
Monolithic integration supports tailoring of individual layers for specific optical properties such as dispersion, transparency range, or nonlinear response while maintaining lithographic alignment and low-loss interlayer transitions.
By preserving full wafer-scale compatibility and avoiding intermediate assembly steps, monolithic integration offers a framework to combine high-performance nonlinear functionality with complementary active and passive photonic devices.

Here, we introduce monolithic 3D integration of tantala integrated photonics on arbitrary substrates, leveraging ultra-low loss across the visible and near-IR plus nonlinear and nanophotonic functionality.
Tantala, sputtered in thick films directly onto wafer substrates, affords the advantages of direct deposition while also presenting a suite of high-performance characteristics as a broadband-low-loss visible and near-IR $\chi^{(3)}$ nonlinear platform \cite{jung_tantala_2021}.
Further, owing to its low residual stress and low temperature annealing requirement, tantala presents the opportunity for nearly universal compatibility in integration with underlying photonic materials and structures.
To demonstrate seamless integration, we realize a tantala-LN platform, choosing thin-film LN on silicon as the substrate for its appeal as a $\chi^{(2)}$ and electro-optic visible photonics platform that affords novel nonlinear operation when paired with $\chi^{(3)}$ tantala.
We present ultra-low loss, 3D integrated circuits of tantala-LN, demonstrating interlayer routing that enables upper-layer tantala photonics to efficiently interface lower-layer LN devices.
We verify low-loss waveguide fabrication by measuring microresonators with intrinsic quality factors ($Q_i$) exceeding 5 million and characterize high-Q photonic crystal resonators (PhCRs) \cite{lu2014selective,yu2021spontaneous,black_optical-parametric_2022,zang_foundry_2024,jin_bandgap-detuned_2025}, demonstrating platform support for nanophotonic designs.
Finally, we showcase $\chi^{(3)}$ and $\chi^{(2)}$ tantala-LN operation with four-wave-mixing (FWM) optical parametric oscillation (OPO) and soliton microcomb generation on upper-layer tantala, second-harmonic generation in lower-layer periodically poled LN, and the combined operation of FWM with SHG in a series connected $\chi^{(3)}$-$\chi^{(2)}$ device. Our results demonstrate monolithic 3D tantala integration as a remarkable new paradigm for scalable, multifunctional photonic systems, enabling incorporation of nonlinear optics into existing photonics infrastructure.

\section{Monolithic 3D Integration of Tantala}
\begin{figure*}[t!]
    \centering
    \includegraphics[width=\linewidth]{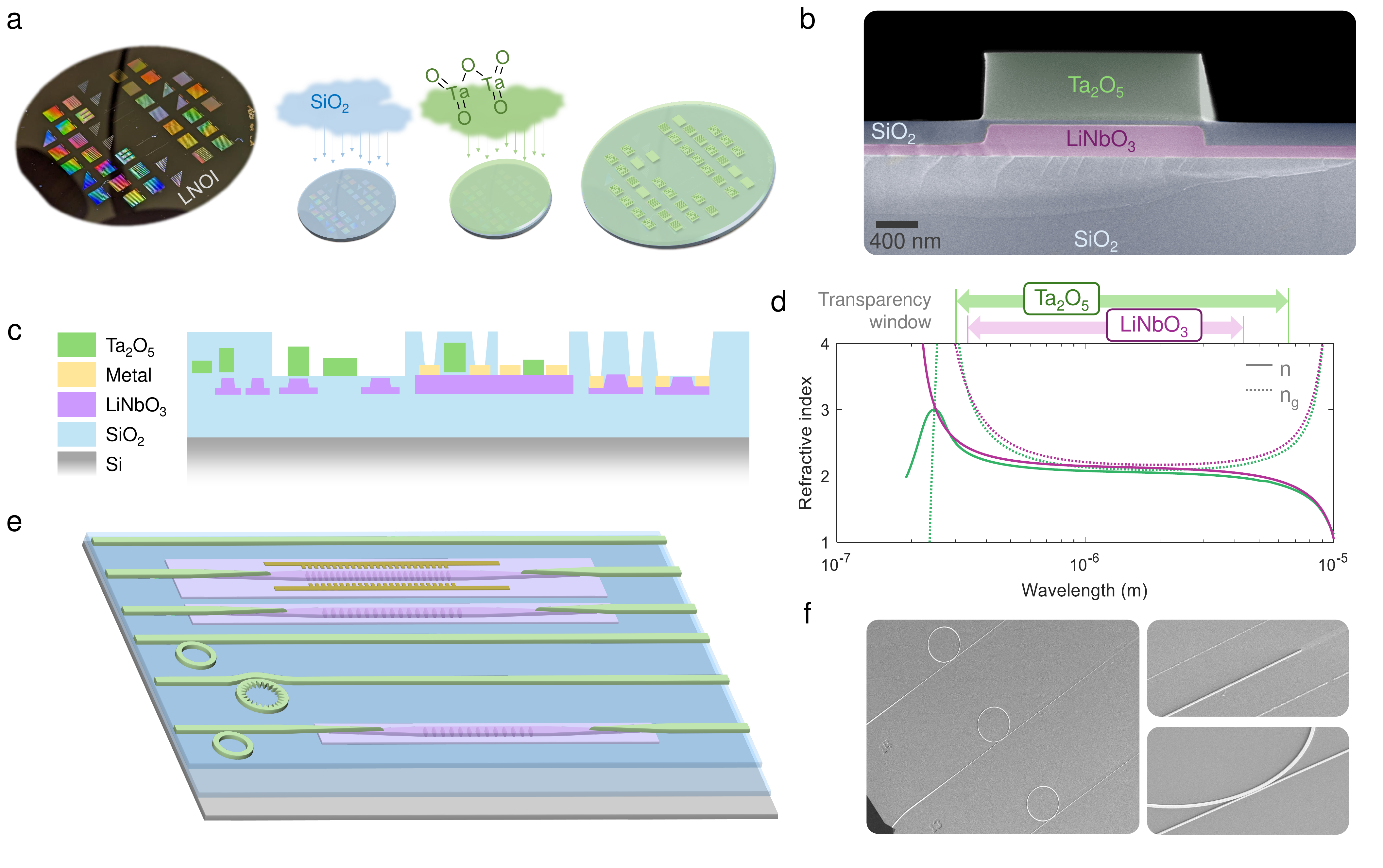}
    \caption{(a) Direct deposition of tantala onto arbitrary substrates, functionalizing photonic platforms using monolithic 3D processing. (b) Scanning electron microscope (SEM) of tantala-LN cross section. (c) Tantala-LN waveguide configurations and metal integration. (d) Bulk material and group indices of refraction for tantala \cite{black_group-velocity-dispersion_2021} and lithium niobate \cite{gayer_temperature_2008}, with transparency windows spanning visible to SWIR identified. (e) Tantala-LN device palette comprising poled LN waveguides, high-Q tantala microresonators, and photonic crystal resonators with 3D interlayer routing afforded by taper structures. (f) SEM images of tantala resonators (left), an interlayer taper transition between tantala and LN (right, top), and a tantala ring-bus coupler (right, bottom). }
    \label{fig:fig1}
\end{figure*}

Figure 1 presents our monolithic 3D integration approach for realizing tantala photonics on arbitrary substrates. Our process employs room temperature ion-beam sputtering (IBS) of tantala to leverage a mix of capabilities for monolithic integration, such as precise control of tantala films onto substrates and amenability to doping with, e.g., metal-oxide mixtures to tailor material properties \cite{carollo_amorphous_2025}. Figure 1a explores the flexibility of monolithic 3D integration with thin-film LN-on-insulator (LNOI) as the lower substrate, realizing the tantala-LN platform. Our process begins with patterning the LN layer, followed by inductively coupled plasma chemical vapor deposition (ICPCVD) silicon-oxide deposition and planarization, then IBS and patterning of the tantala layer. The ultrathin, planarized oxide layer enables a low-loss interface between tantala photonics with the lower substrate.
Figure 1b shows a cross section scanning electron microscope (SEM) image of an example structure in which a tantala waveguide is realized directly on top of an etched LN waveguide.
In this image, the bottom oxide and LN layer are components of the LNOI wafer, while the upper oxide is the planarized interface layer. 
Here, the LN layer is partially etched to realize a specific device design, but full layer etching is also possible. The upper layer tantala waveguides shown in the SEM are designed for air-cladding to realize high-index-contrast operation for the visible wavelength band. This image qualitatively assesses waveguide fabrication and alignment precision realized by monolithic tantala processing, thereby demonstrating capabilities for direct, co-integration of two distinct waveguide layers.

In Fig. 1c, we show the range of waveguide designs supported by monolithic 3D tantala-LN integration in terms of material layers, claddings, and optional metallization for poling electrodes, electro-optic control, and heaters. The platform leverages the complementary optical properties of tantala and LN (Fig. 1d), specifically, broadband transparency from the upper edge of the ultraviolet to SWIR and comparable refractive indices that support high index contrast waveguides and efficient mode matching between waveguides \cite{black_group-velocity-dispersion_2021,gayer_temperature_2008}. Together, the material properties and the layout circuit possibilities create the functionality to access an exceptionally rich nonlinear device design space (Fig. 1e). Figure 1f shows SEM images of a tantala microresonator routed to lower layer LN waveguides through an interlayer taper structure. Inverse taper transitions in both the LN and tantala layers offers access to arbitrary circuits of the materials.

\section{Fabrication and passive characteristics}
\begin{figure*}[t!]
    \centering
    \includegraphics[width=\linewidth]{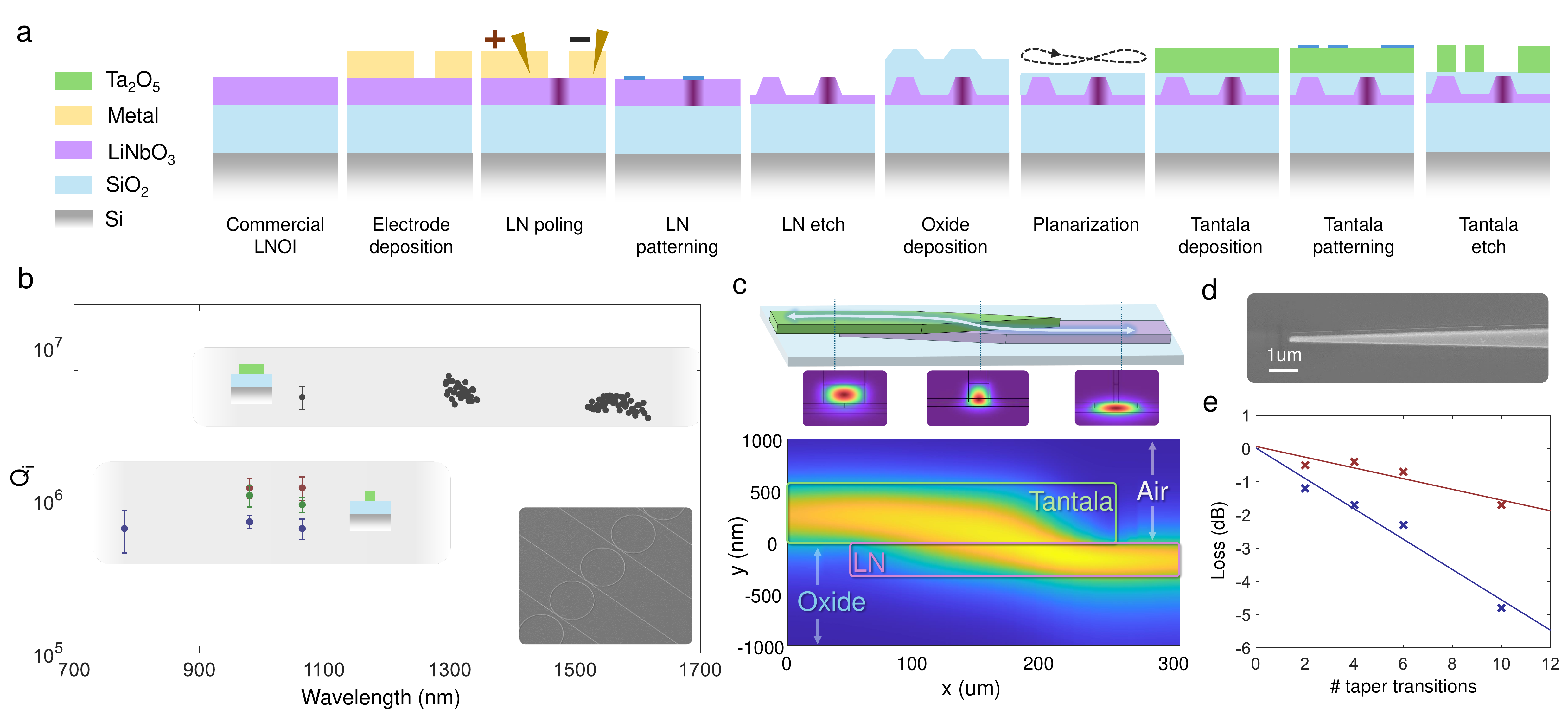}
    \caption{(a) Process flow for monolithic 3D integrated tantala photonics realized directly on LNOI substrates. (b) Measured intrinsic quality factor ($Q_i$) of wide ring width ($RW$), normal dispersion (top, $RW$ = 4 um) and narrow $RW$ near-zero dispersion (bottom, $RW$ = 890, 1040, and 1140 um in blue, green, and red, respectively)  microresonators fabricated on the tantala-LN platform. (c) Interlayer routing taper schematic, simulated mode profiles at points in the taper transition, and 3D modeled E-field showing transmission from upper layer tantala to lower layer LN. (d) Scanning electron microscope image of a fabricated tantala taper. (e) Measured loss of taper transition test structures at 1550 nm (red) and 780 nm (blue).}
    \label{fig:fig2}
\end{figure*}

We outline the tantala-LN wafer fabrication process in Fig. 2a. 
Processing begins with a commercial, 300 nm thick magnesium-oxide-doped X-cut lithium niobate film on a three inch oxidized silicon wafer. 
Devices on the LN layer like periodically poled SHG (PPLN) waveguides are processed prior to tantala deposition. 
To realize electrodes for periodically poling in LN, we use electron beam lithography (EBL) to create a pattern in ZEP520A resist, deposit a 100 nm SiO\textsubscript{2} interlayer via atomic layer deposition, then a 100 nm thick layer of gold-chromium metal.
Removing the EBL resist transfers the pattern to the metal layer via liftoff.
Our electrode designs are electrically linked across neighboring devices to parallelize poling and lend scalability to fabrication.
We create poled domains in the LN layer by use of a voltage pulse waveform that exceeds the coercive field strength of LN \cite{younesi_periodic_2021,rao_actively-monitored_2019}, and we monitor the domain inversion using a two-photon microscope. 
After removing poling electrode metals, we lithographically pattern and etch LN waveguides, using an argon ion mill to a partial etch depth of approximately 150 nm. To prepare the lower substrate surface for tantala deposition, we proceed by depositing oxide by use of ICPCVD. Because the oxide inherits the topology of the devices and structures present on the lower substrate, we use chemical-mechanical polishing (CMP) to planarize the wafer and prepare it for tantala deposition. Following CMP, the substrate resembles a standard, oxide-clad wafer from the perspective of tantala processing, allowing for fabrication to proceed with no modifications from previously developed process flows.
We deposit tantala, in this work to a thickness of 570 nm, using room temperature IBS. 
We use EBL to define tantala structures on an 85 nm thick alumina hardmask layer, followed by etching in a fluorene-based plasma and mask removal.
To reduce photorefractive effects in tantala \cite{fazio_structure_2020}, we deposit a 20 nm thick, conformal layer of SiO\textsubscript{2} via atomic-layer deposition on the tantala layer and anneal the wafer at 500 $^{\circ}$C for 12 hours (this anneal step is bypassed for titanium-doped tantala \cite{carollo_amorphous_2025}).
Finally, tantala-LN wafers are singulated into individual chips using a deep-reactive-ion-etch (DRIE) which optionally allows for angled facets and other facet geometries.

We evaluate tantala-LN passive optical characteristics by measuring waveguides, resonators, and 3D interlayer routing. 
For microresonator-based FWM, high Q is of key importance, since threshold power scales inversely proportional to the square of loaded Q \cite{yi_soliton_2015,lu_milliwatt-threshold_2019} and integrated devices aim for operation with practical on-chip pump powers.
We measure the intrinsic Q of dispersion-engineered, wide and narrow ring width ($RW$) microresonators in Fig. 2b.
Experimentally, we perform microresonator spectroscopy at multiple wavelengths with a suite of external cavity tunable lasers.
We tune each laser across device resonances in parallel with a calibrated fibre Mach-Zehnder interferometer (MZI), employing the MZI transmission response as an optical frequency calibration.
The devices we characterize operate on the fundamental transverse electric mode family (TE0).
In devices with a $RW$ of 4 um, optimized for normal dispersion 200 GHz dark pulse generation, we measure intrinsic quality factors on the order of 5 Million at wavelengths spanning 1064 nm to 1600 nm; see Fig. 2b, top panel.
In narrow $RW$ devices (Fig. 2b, bottom) designed for near-zero dispersion broadband OPO, we measure $Q_i$ of $\sim7\times10^5$ for 890 nm $RW$ (blue), $\sim9\times10^5$ for 1040 nm $RW$ (green), and $\sim1\times10^6$ for 1140 $RW$ (red) at wavelengths spanning 780 nm to 1064 nm.
The dependence of $Q_i$ on $RW$ supports the conclusion that optical loss is limited by waveguide scattering due to increased modal overlap with sidewalls in narrower devices \cite{roberts_measurements_2022}. 

With confirmation of low-loss tantala waveguides, the next critical performance aspect for operation of 3D tantala-LN circuits is efficient routing of light between the upper and lower layers.
To realize broadband, low-loss interlayer transitions, we design 3D structures comprising vertically coupled, collinear inverse tapers, whereby the optical mode is adiabatically transferred from one waveguide layer to the other. The taper regions are 250 $\mu$m in total length, comprising a linearly decreasing waveguide width from 2 $\mu$m down to 150 nm for both the tantala and LN waveguides.
We show a 3D schematic (exaggerated dimensions to show features) and finite-difference-time-domain (FDTD) simulations of transition structures in Fig. 2c.
In Fig. 2d, we qualitatively inspect the smallest feature size of a 150 nm taper tip, using an SEM to confirm fidelity of our designs realized on fabricated wafers.
To experimentally characterize interlayer routing loss, we design test devices comprising serially cascaded taper structures with a varying number of transitions across devices.
We measure loss-per-transition below $\sim$0.2 dB and $\sim$0.5 dB for test wavelengths at 1550 nm (red) and 780 nm (blue), respectively; see Fig. 2e.

\section{Nonlinear operation}
In this section, we explore design and operation of tantala-LN devices, demonstrating a complete palette of ultra-efficient $\chi^{(2)}$ and $\chi^{(3)}$ nonlinear optics and circuit combinations.
We explore SHG in $\chi^{(2)}$ PPLN waveguides, especially through generation of visible light, and in 3D tantala-LN devices, where light is routed between tantala and LN layers to demonstrate interlayer functionality.
In $\chi^{(3)}$ tantala, we demonstrate efficient FWM processes that realize ultra-broadband OPO in near-zero dispersion microresonators \cite{brodnik_nanopatterned_2025,domeneguetti_parametric_2021,sayson_octave-spanning_2019} and dark pulse generation in normal dispersion photonic crystal resonators \cite{yu2021spontaneous,zang_laser_2025,jin_bandgap-detuned_2025}.  Finally, we demonstrate cascaded $\chi^{(3)}$-$\chi^{(2)}$ operation by cascading upper-layer tantala microresonator OPO to a poled LN waveguide to frequency double the OPO in a single device.

\begin{figure*}[t!]
\centering
    \includegraphics[width=\linewidth]{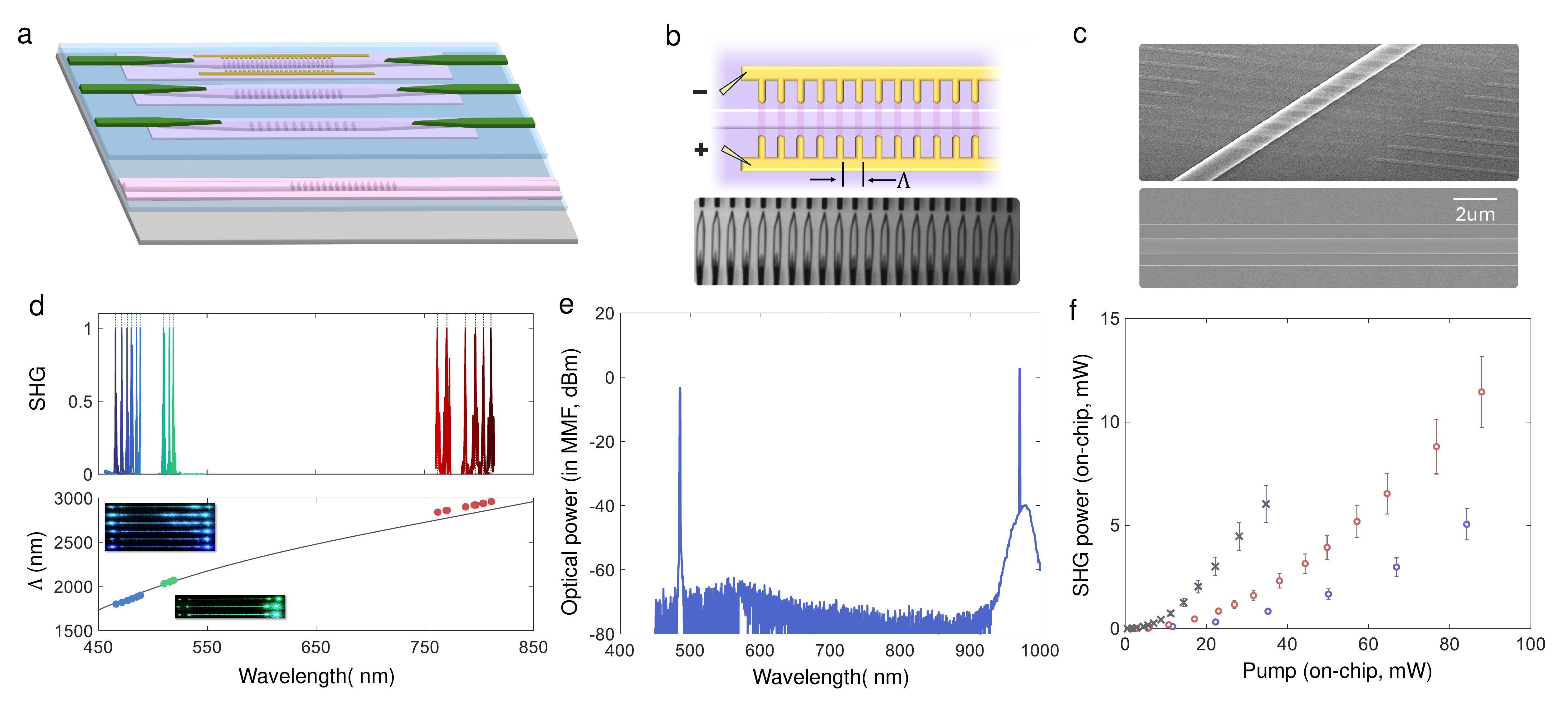}
    \caption{(a) Second harmonic test devices in LNOI and tantala-LN wafers. (b) Schematic of LN poling electrodes (top) and second harmonic microscope image showing domain inversion of poled regions (bottom). (c) Scanning electron microscope images of a poled and etched LN waveguide. (d) Normalized SHG gain spectrum vs. wavelength for devices with varying $\Lambda$ (top). Comparison to phase matching models based on uniform LN thickness, and images of devices during operation (bottom). (e) Optical spectrum of a tantala-LN SHG device converting a 970 nm pump to 485 nm signal, measured directly from multimode fibre edge-coupled to output facet. (f) Calibrated on-chip pump power vs. on-chip SHG power for a LN device (grey x's) and tantala-LN devices with output SHG wavelengths of 485 nm (blue o's) and 787 nm (red o's).}
    \label{fig:fig3}
\end{figure*}

Figure 3 characterizes $\chi^{(2)}$ SHG designs, which depend on poling quality, quasi-phase-matching wavelength accuracy, and SHG efficiency across tantala-LN 3D devices (Fig. 3a, top) and LNOI benchmark devices (Fig. 3a, bottom). In Fig. 3b, we show a schematic of poling electrodes centered about a LN waveguide and define poling pitch, $\Lambda$, that controls QPM and SHG gain wavelength. Throughout the development of the poling procedure, we visually monitor domain inversion by use of a second-harmonic (SH) microscope. In the SH microscope image shown in Fig. 3b (bottom), poling electrodes are seen as dark fringes along the top and bottom, and inverted domains as the segments spanning the middle. Qualitatively, uniform coloring within each individual domain suggests complete poling through LN films while fixed repetition across domains shows longitudinal uniformity \cite{younesi_periodic_2021,rao_actively-monitored_2019}. We show SEMs of poled devices following waveguide etching in Fig. 3c. Residual electrode material is observed on either side of the waveguide, but at distances sufficiently far from the waveguide to avoid interaction with optical modes.

We characterize wavelength-dependent phase matching in devices that target a range of visible SHG wavelengths by varying $\Lambda$ (Fig. 3d). The devices are selected from two separate chips on the same tantala-LN wafer, one chip having devices poled for pump wavelengths spanning $\sim$920-980 nm (blue curves and dots) and the other for $\sim$1020-1050 nm (green curves and dots). Experimentally, we couple pump light into the chip using a lensed fibre with spot size of 2 $\mu$m, and collect SHG light out of the chip by edge coupling a multimode fibre directly to the chip facet. We tune the wavelength of the pump laser across the range of designed QPM targets and use an optical power meter with short pass dichroic filtering, cut at 750 nm, to record SHG power versus wavelength. We plot the measured SHG gain spectrum for each $\Lambda$ design (solid lines) and the calculated gain centers of mass (dotted lines) in the top of Fig. 3d. In the bottom of Fig. 3d, we plot the extracted gain centers of mass for each $\Lambda$ along with the theoretical phase matching curve (grey line).
We note a slight offset in the measured SHG phase matching compared to the modeled curve when comparing the two chips, suggesting wafer-level index deviations. Such effects have been explored in the literature and have been attributed to thickness variations in LN films, which could be compensated in future fabrication runs by adaptively biasing $\Lambda$ designs based on wafer level thickness measurements \cite{chen_adapted_2024,xin_wavelength-accurate_2025}.

We next characterize SHG devices operating under high-power pump conditions to assess nonlinear performance in terms of power and efficiency. 
We show a sample spectrum of a 485 nm SHG device in Fig. 3e.
In Fig. 3f, we plot the calibrated on-chip pump power vs. SHG power of a LN device designed for an SHG target of 550 nm with device length, $L$, of 2000 um (grey x's), and of tantala-LN devices for SHG targets of 485 nm (blue o's) and 787 nm (red o's), both with $L$ = 1800 um. 
We measure on-chip SHG power exceeding 6 mW for an on-chip pump power of 35 mW in the LN device, SHG power exceeding 5 mW for a pump power of 84 mW in the 485 nm tantala-LN device, and SHG power of 11 mW for a pump power of 87 mW in the 787 nm tantala-LN device.
We calculate the geometry-normalized conversion efficiency to be 13000~\%$\textrm{W}^{-1}\textrm{cm}^{-2}$ for the LN device, 2200~$\%\textrm{W}^{-1} \textrm{cm}^{-2}$ for the 485 nm tantala-LN device, and 4500~$\%\textrm{W}^{-1} \textrm{cm}^{-2}$ for the 787 nm tantala-LN device.

Figure 4 explores $\chi^{(3)}$ and paired $\chi^{(3)}$-$\chi^{(2)}$ devices in monolithically integrated tantala-LN, which draw on the transition of previously developed tantala microresonator and photonic crystal resonator device designs \cite{yu2021spontaneous,black_optical-parametric_2022,zang_foundry_2024,jin_bandgap-detuned_2025,brodnik_nanopatterned_2025}.
In $\chi^{(3)}$ devices realized in upper layer air-clad tantala, we employ 570 nm thick films that support strong mode confinement for high nonlinear gain and designable dispersion for controlling nonlinear phase matching. We characterize geometric control of phase-matching by simulating group-velocity dispersion (GVD) for varied microresonator waveguide geometries that realize normal, near-zero, and anomalous GVD across near-IR pump laser frequencies; see Fig. 4a.
To link geometric dispersion engineering with practical designs, we identify the near-zero dispersion crossing (dotted lines) for each $RW$, highlighting the operating point that realizes ultra-broadband wavelength conversion based on OPO. 

\begin{figure*}[t!]
    \centering
    \includegraphics[width=\linewidth]{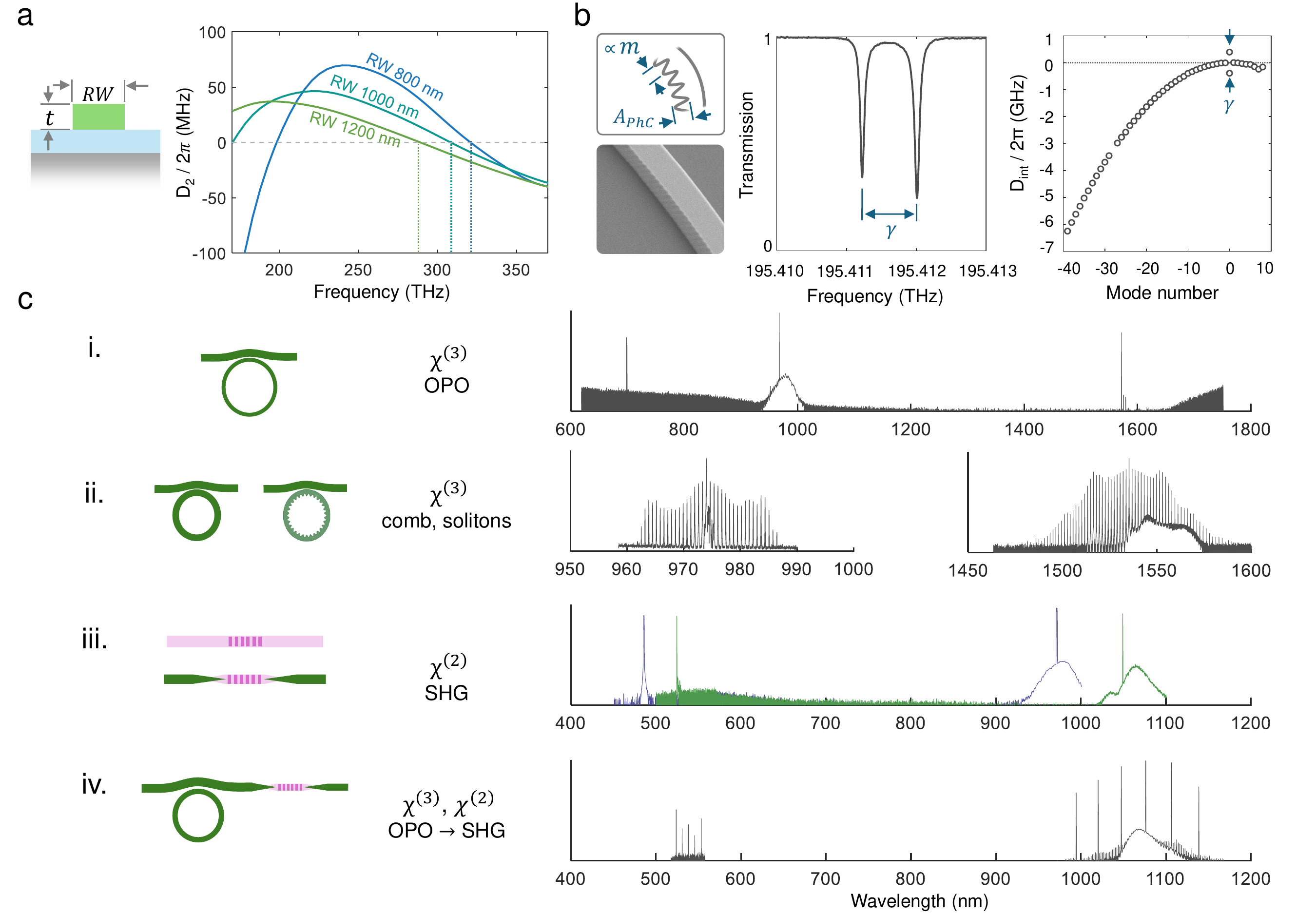}
    \caption{(a) Geometric dispersion engineering and control of $D_2$ GVD parameter, realized by designing waveguide thickness, $t$, and ring width, $RW$. (b) Photonic crystal design showing SEM image (left), measured resonator mode split, $\gamma$  (middle), and integrated dispersion of a negative $D_2$ device with photonic crystal on the pump mode (right). (c) Summary of nonlinear operation realized in 3D tantala-LN photonics: (i) broadband wavelength conversion in near-zero dispersion tantala microresonator OPO, (ii) dark soliton generation in normal dispersion tantala photonic crystal resonators, (iii) second harmonic generation in 3D tantala-LN using poled LN waveguides, and (iv) cascaded $\chi^{(3)}$ OPO in tantala layer to $\chi^{(2)}$ SHG in lower LN layer.}
    \label{fig:fig4}
\end{figure*}

Next, we show nanophotonic designs that leverage nanometer-scale resolution etching of tantala to fabricate high-Q PhCRs \cite{lu2014selective,yu_continuum_2022,yu2021spontaneous}.
Such designs control the PhCR bandgap mode and magnitude to lend mode-programmable control of phase matching. We present a photonic crystal design realized via microresonator sidewall modulations in Fig. 4b, defining $m$ as the parameter that affords mode selectivity via Bragg matching and $A_{PhC}$ as the parameter that controls bandgap magnitude, $\gamma$.
We show an SEM of a fabricated photonic crystal device in Fig. 4b (bottom left).
We measure $\gamma$ in Fig. 4b (middle), and the integrated dispersion, D\textsubscript{int} \cite{fujii2020dispersion}(right).
The combination of normal GVD, realized through design of $RW$, and phase-matching for FWM, realized through design of the pump mode bandgap, enables dark soliton microcomb generation \cite{yu_continuum_2022}.

In Fig. 4c, we present experimental measurements that characterize the complete palette of $\chi^{(3)}$ and $\chi^{(2)}$ nonlinear photonics on the tantala-LN platform. 
Our focus is to explore the breadth of devices that can be realized by the scale of even 3 inch wafer processing. 
These experiments encompass major themes in the recent development of integrated nonlinear photonics for the generation of unique laser sources, including soliton microcombs, SHG, and broadband OPOs. We schematically show tantala-LN circuits and their measured output optical spectra. 
In a tantala microresonator designed for near-zero dispersion via geometric control (Fig. 4a), we demonstrate wavelength conversion that exceeds an octave of optical bandwidth from a pump wavelength of 968 nm to signal and idler wavelengths of 698 nm, and 1572 nm, respectively (Fig. 4c(i)). 
In a normal dispersion microresonator, and the normal dispersion PhCR (Fig. 4b), we show dark pulse generation at pump wavelengths of 974 nm and 1535 nm, respectively (Fig. 4c(ii)). In Fig. 4c(iii), we show second harmonic generation in LN waveguides poled for 1050-525 nm SHG (plotted in green), and in tantala-LN devices for 970-485 SHG (plotted in blue). 
Finally, Fig. 4c(iv) shows upper layer $\chi^{(3)}$ OPO in tantala routed to lower layer $\chi^{(2)}$ LN for SHG in a single tantala-LN device. 
Here, the pump laser is at 1076 nm, making primary OPO signal and idler waves at 1048 nm and 1107 nm, respectively. 
The full OPO output is routed from the upper layer tantala waveguide into the lower layer LN waveguide, with QPM peak SHG gain measured at 524 nm. 
We observe the highest conversion of the 1048 nm to 524 nm wave, as it aligned best with the QPM peak.
We also observe weaker SHG corresponding to conversion of the residual pump laser from 1076 nm to 538 nm, of the idler from 1107 nm to 553 nm, and additionally sum-frequency generation (SFG) of pump-plus-signal and pump-plus-idler OPO waves.

\section{Conclusion}
We have presented monolithic 3D integration that combines tantala photonics with arbitrary substrates, exploring thin-film lithium niobate in this work. Our approach uses room-temperature deposition and low-temperature processing of tantala, enabling broad compatibility with existing photonic materials and foundry infrastructure. We achieve low-loss tantala waveguides supporting high-$Q$ microresonators and nanophotonic devices, efficient SHG through periodic poling in LN, and low-loss interlayer routing that bridges the two materials. To highlight nonlinear functionality, we demonstrate $\chi^{(3)}$ four-wave mixing and soliton microcomb generation in tantala, $\chi^{(2)}$ second harmonic generation in LN, and cascaded $\chi^{(3)}$–$\chi^{(2)}$ operation in a unified device. These results establish monolithic 3D tantala integration as a scalable and versatile platform for visible and near-IR photonics, uniting complementary nonlinear materials to enable compact, multifunctional, and manufacturable photonic systems.

\section{Methods}

\textbf{Second-harmonic quasi-phase-matching}
\newline We model the QPM curve in the bottom of Fig. 3d by first solving for the wavelength-dependent effective index of the fundamental TE mode across the range of pump and SHG wavelengths using a finite element method (FEM) mode solver. For material refractive indices, we use Sellmeier data for 5\% MgO-doped LN \cite{gayer_temperature_2008} and SiO$_2$ \cite{malitson_interspecimen_1965}. The waveguide design comprises a SiO$_2$ cladding with a 300 nm thick LN waveguide layer, partially etched a depth of 150 nm to realize a 2 $\mu$m wide ridge. We use the FEM wavelength dependent effective index data to link pump and SHG wavevectors with $\Lambda$ using the quasi-phase-matching relation \cite{boyd_nonlinear_optics},
\begin{equation}\label{eq:qpm}
    \Lambda = 2\pi/(2k_{\omega}-k_{2\omega}).
\end{equation}

\vspace{5mm}
\noindent\textbf{Second harmonic conversion efficiency}
\newline We calculate SHG conversion efficiency by measuring output SHG power versus input pump power (Fig. 3f). Pump light is coupled into test devices using HI1060 lensed fibre with a 2 $\mu$m spot size, and SHG light is collected by edge coupling a 50 $\mu$m core diameter multimode fibre directly to the output facet. We calibrate insertion loss arising from fibre-to-chip coupling using a tantala straight waveguide test structure, coupling into and out of the chip using lensed fibre of the same design, and assume facet loss is symmetric. We then replace the output fibre with the multimode fibre and measure the additional loss to calibrate insertion loss for the output. We fit the SHG power versus pump data in Fig. 3f to a quadratic curve, assuming no pump depletion, to calculate conversion efficiency in terms of $\%\textrm{W}^{-1}$, then normalize to units of $\%\textrm{W}^{-1}\textrm{cm}^{-2}$ using the device lengths reported in text.

\vspace{5mm}
\noindent\textbf{Tantala dispersion design}
\newline We model group velocity dispersion in Fig. 4a by first solving for the effective index as a function of wavelength of the TE0 mode using an FEM mode solver. For the material refractive index of tantala, we use a Sellmeier fit from previously published ellipsometry data measured on the tantala photonics platform \cite{black_group-velocity-dispersion_2021}. We use the calculated effective index data to solve for discrete resonator mode frequencies, $\omega$, with corresponding integer mode numbers, $m$, for a given microresonator ring radius. For a selected pump mode, $m_0$, we then calculate integrated dispersion using
\begin{equation} \label{dint_polynomial}
\begin{split}
 \omega(\mu) & =  \omega_{0} + {D_1}\mu + \frac{D_2\mu^2}{2} + \frac{D_3\mu^3}{6} + ... \\
 & = {\omega_0} + {D_1}\mu + D_\textrm{int}.
\end{split}
\end{equation}
where $\mu$ is the mode number with respect to the pump ($\mu = m-m_0$).
In this description, $D_1/{2\pi}$ is the free spectral range (FSR). $D_{int}$ thereby describes the departure of mode frequencies from a linear, FSR-spaced grid. $D_2/{2\pi}$ is the GVD parameter plotted in Fig. 4a. 

\section*{Acknowledgments}
We thank Nitesh Chauhan and Sarang Yeola for careful review of the manuscript.
\paragraph*{Funding:}
This research has been funded by the DARPA LUMOS program HR0011-20-2-0046, AFOSR FA9550-20-1-0004 Project Number 19RT1019, NSF Quantum Leap Challenge Institute Award OMA - 2016244, and NIST.
\paragraph*{Competing interests:}
The authors declare no competing interests. This work is a contribution of the US Government and is not subject to US copyright. Mention of specific companies or trade names is for scientific communication only and does not constitute an endorsement by NIST. 
\paragraph*{Data and materials availability:}
All data necessary to evaluate the conclusions of this work is provided in the paper.

\clearpage
\bibliography{Tantala-LN_BIB}

\begin{thebibliography}{10}
\expandafter\ifx\csname url\endcsname\relax
  \def\url#1{\burl{#1}}\fi
\expandafter\ifx\csname urlprefix\endcsname\relax\def\urlprefix{URL }\fi
\providecommand{\bibinfo}[2]{#2}
\providecommand{\eprint}[2][]{\url{#2}}
\providecommand{\doi}[1]{\url{https://doi.org/#1}}

\bibitem{spencer_optical-frequency_2018}
\bibinfo{author}{Spencer, D.~T.} \emph{et~al.}
\newblock \bibinfo{title}{An optical-frequency synthesizer using integrated photonics}.
\newblock \emph{\bibinfo{journal}{Nature}} \textbf{\bibinfo{volume}{557}}, \bibinfo{pages}{81--85} (\bibinfo{year}{2018}).

\bibitem{pfeifle_coherent_2014}
\bibinfo{author}{Pfeifle, J.} \emph{et~al.}
\newblock \bibinfo{title}{Coherent terabit communications with microresonator {Kerr} frequency combs}.
\newblock \emph{\bibinfo{journal}{Nature Photonics}} \textbf{\bibinfo{volume}{8}}, \bibinfo{pages}{375--380} (\bibinfo{year}{2014}).

\bibitem{raval_integrated_2018}
\bibinfo{author}{Raval, M.}, \bibinfo{author}{Yaacobi, A.} \& \bibinfo{author}{Watts, M.~R.}
\newblock \bibinfo{title}{Integrated visible light phased array system for autostereoscopic image projection}.
\newblock \emph{\bibinfo{journal}{Optics Letters}} \textbf{\bibinfo{volume}{43}}, \bibinfo{pages}{3678--3681} (\bibinfo{year}{2018}).

\bibitem{alexander_manufacturable_2025}
\bibinfo{author}{Alexander, K.} \emph{et~al.}
\newblock \bibinfo{title}{A manufacturable platform for photonic quantum computing}.
\newblock \emph{\bibinfo{journal}{Nature}} \textbf{\bibinfo{volume}{641}}, \bibinfo{pages}{876--883} (\bibinfo{year}{2025}).

\bibitem{pfeiffer_photonic_2016}
\bibinfo{author}{Pfeiffer, M. H.~P.} \emph{et~al.}
\newblock \bibinfo{title}{Photonic {Damascene} process for integrated high-{Q} microresonator based nonlinear photonics}.
\newblock \emph{\bibinfo{journal}{Optica}} \textbf{\bibinfo{volume}{3}}, \bibinfo{pages}{20--25} (\bibinfo{year}{2016}).

\bibitem{jung_tantala_2021}
\bibinfo{author}{Jung, H.} \emph{et~al.}
\newblock \bibinfo{title}{Tantala kerr nonlinear integrated photonics}.
\newblock \emph{\bibinfo{journal}{Optica}} \textbf{\bibinfo{volume}{8}}, \bibinfo{pages}{811--817} (\bibinfo{year}{2021}).

\bibitem{jung_aluminum_2016}
\bibinfo{author}{Jung, H.} \& \bibinfo{author}{Tang, H.~X.}
\newblock \bibinfo{title}{Aluminum nitride as nonlinear optical material for on-chip frequency comb generation and frequency conversion}.
\newblock \emph{\bibinfo{journal}{Nanophotonics}} \textbf{\bibinfo{volume}{5}}, \bibinfo{pages}{263--271} (\bibinfo{year}{2016}).

\bibitem{desiatov_ultra-low-loss_2019}
\bibinfo{author}{Desiatov, B.}, \bibinfo{author}{Shams-Ansari, A.}, \bibinfo{author}{Zhang, M.}, \bibinfo{author}{Wang, C.} \& \bibinfo{author}{Lončar, M.}
\newblock \bibinfo{title}{Ultra-low-loss integrated visible photonics using thin-film lithium niobate}.
\newblock \emph{\bibinfo{journal}{Optica}} \textbf{\bibinfo{volume}{6}}, \bibinfo{pages}{380--384} (\bibinfo{year}{2019}).

\bibitem{zhu_integrated_2021}
\bibinfo{author}{Zhu, D.} \emph{et~al.}
\newblock \bibinfo{title}{Integrated photonics on thin-film lithium niobate}.
\newblock \emph{\bibinfo{journal}{Advances in Optics and Photonics}} \textbf{\bibinfo{volume}{13}}, \bibinfo{pages}{242--352} (\bibinfo{year}{2021}).

\bibitem{boes_lithium_2023}
\bibinfo{author}{Boes, A.} \emph{et~al.}
\newblock \bibinfo{title}{Lithium niobate photonics: {Unlocking} the electromagnetic spectrum}.
\newblock \emph{\bibinfo{journal}{Science}} \textbf{\bibinfo{volume}{379}} (\bibinfo{year}{2023}).

\bibitem{wang_lithium_2024}
\bibinfo{author}{Wang, C.} \emph{et~al.}
\newblock \bibinfo{title}{Lithium tantalate photonic integrated circuits for volume manufacturing}.
\newblock \emph{\bibinfo{journal}{Nature}} \textbf{\bibinfo{volume}{629}}, \bibinfo{pages}{784--790} (\bibinfo{year}{2024}).

\bibitem{karvounis_barium_2020}
\bibinfo{author}{Karvounis, A.}, \bibinfo{author}{Timpu, F.}, \bibinfo{author}{Vogler-Neuling, V.~V.}, \bibinfo{author}{Savo, R.} \& \bibinfo{author}{Grange, R.}
\newblock \bibinfo{title}{Barium {Titanate} {Nanostructures} and {Thin} {Films} for {Photonics}}.
\newblock \emph{\bibinfo{journal}{Advanced Optical Materials}} \textbf{\bibinfo{volume}{8}}, \bibinfo{pages}{2001249} (\bibinfo{year}{2020}).

\bibitem{wang_integrated_2018}
\bibinfo{author}{Wang, C.} \emph{et~al.}
\newblock \bibinfo{title}{Integrated lithium niobate electro-optic modulators operating at {CMOS}-compatible voltages}.
\newblock \emph{\bibinfo{journal}{Nature}} \textbf{\bibinfo{volume}{562}}, \bibinfo{pages}{101--104} (\bibinfo{year}{2018}).

\bibitem{wang_ultrahigh-efficiency_2018}
\bibinfo{author}{Wang, C.} \emph{et~al.}
\newblock \bibinfo{title}{Ultrahigh-efficiency wavelength conversion in nanophotonic periodically poled lithium niobate waveguides}.
\newblock \emph{\bibinfo{journal}{Optica}} \textbf{\bibinfo{volume}{5}}, \bibinfo{pages}{1438--1441} (\bibinfo{year}{2018}).

\bibitem{jankowski_ultrabroadband_2020}
\bibinfo{author}{Jankowski, M.} \emph{et~al.}
\newblock \bibinfo{title}{Ultrabroadband nonlinear optics in nanophotonic periodically poled lithium niobate waveguides}.
\newblock \emph{\bibinfo{journal}{Optica}} \textbf{\bibinfo{volume}{7}}, \bibinfo{pages}{40--46} (\bibinfo{year}{2020}).

\bibitem{younesi_periodic_2021}
\bibinfo{author}{Younesi, M.} \emph{et~al.}
\newblock \bibinfo{title}{Periodic poling with a micrometer-range period in thin-film lithium niobate on insulator}.
\newblock \emph{\bibinfo{journal}{JOSA B}} \textbf{\bibinfo{volume}{38}}, \bibinfo{pages}{685--691} (\bibinfo{year}{2021}).

\bibitem{park_high-efficiency_2022}
\bibinfo{author}{Park, T.} \emph{et~al.}
\newblock \bibinfo{title}{High-efficiency second harmonic generation of blue light on thin-film lithium niobate}.
\newblock \emph{\bibinfo{journal}{Optics Letters}} \textbf{\bibinfo{volume}{47}}, \bibinfo{pages}{2706--2709} (\bibinfo{year}{2022}).

\bibitem{komljenovic_photonic_2018}
\bibinfo{author}{Komljenovic, T.} \emph{et~al.}
\newblock \bibinfo{title}{Photonic {Integrated} {Circuits} {Using} {Heterogeneous} {Integration} on {Silicon}}.
\newblock \emph{\bibinfo{journal}{Proceedings of the IEEE}} \textbf{\bibinfo{volume}{106}}, \bibinfo{pages}{2246--2257} (\bibinfo{year}{2018}).

\bibitem{fathpour_heterogeneous_2018}
\bibinfo{author}{Fathpour, S.}
\newblock \bibinfo{title}{Heterogeneous {Nonlinear} {Integrated} {Photonics}}.
\newblock \emph{\bibinfo{journal}{IEEE Journal of Quantum Electronics}} \textbf{\bibinfo{volume}{54}}, \bibinfo{pages}{1--16} (\bibinfo{year}{2018}).

\bibitem{nader_heterogeneous_2025}
\bibinfo{author}{Nader, N.} \emph{et~al.}
\newblock \bibinfo{title}{Heterogeneous tantala photonic integrated circuits for sub-micron wavelength applications}.
\newblock \emph{\bibinfo{journal}{Optica}} \textbf{\bibinfo{volume}{12}}, \bibinfo{pages}{585--593} (\bibinfo{year}{2025}).

\bibitem{beeck_iiiv--lithium_2021}
\bibinfo{author}{Beeck, C. O.~d.} \emph{et~al.}
\newblock \bibinfo{title}{{III}/{V}-on-lithium niobate amplifiers and lasers}.
\newblock \emph{\bibinfo{journal}{Optica}} \textbf{\bibinfo{volume}{8}}, \bibinfo{pages}{1288--1289} (\bibinfo{year}{2021}).

\bibitem{snigirev_ultrafast_2023}
\bibinfo{author}{Snigirev, V.} \emph{et~al.}
\newblock \bibinfo{title}{Ultrafast tunable lasers using lithium niobate integrated photonics}.
\newblock \emph{\bibinfo{journal}{Nature}} \textbf{\bibinfo{volume}{615}}, \bibinfo{pages}{411--417} (\bibinfo{year}{2023}).

\bibitem{xiang_laser_2021}
\bibinfo{author}{Xiang, C.} \emph{et~al.}
\newblock \bibinfo{title}{Laser soliton microcombs heterogeneously integrated on silicon}.
\newblock \emph{\bibinfo{journal}{Science}} \textbf{\bibinfo{volume}{373}}, \bibinfo{pages}{99--103} (\bibinfo{year}{2021}).

\bibitem{vanackere_heterogeneous_2023}
\bibinfo{author}{Vanackere, T.} \emph{et~al.}
\newblock \bibinfo{title}{Heterogeneous integration of a high-speed lithium niobate modulator on silicon nitride using micro-transfer printing}.
\newblock \emph{\bibinfo{journal}{APL Photonics}} \textbf{\bibinfo{volume}{8}}, \bibinfo{pages}{086102} (\bibinfo{year}{2023}).

\bibitem{chang_heterogeneous_2017}
\bibinfo{author}{Chang, L.} \emph{et~al.}
\newblock \bibinfo{title}{Heterogeneous integration of lithium niobate and silicon nitride waveguides for wafer-scale photonic integrated circuits on silicon}.
\newblock \emph{\bibinfo{journal}{Optics Letters}} \textbf{\bibinfo{volume}{42}}, \bibinfo{pages}{803--806} (\bibinfo{year}{2017}).

\bibitem{ghosh_wafer-scale_2023}
\bibinfo{author}{Ghosh, S.} \emph{et~al.}
\newblock \bibinfo{title}{Wafer-scale heterogeneous integration of thin film lithium niobate on silicon-nitride photonic integrated circuits with low loss bonding interfaces}.
\newblock \emph{\bibinfo{journal}{Optics Express}} \textbf{\bibinfo{volume}{31}}, \bibinfo{pages}{12005--12015} (\bibinfo{year}{2023}).

\bibitem{churaev_heterogeneously_2023}
\bibinfo{author}{Churaev, M.} \emph{et~al.}
\newblock \bibinfo{title}{A heterogeneously integrated lithium niobate-on-silicon nitride photonic platform}.
\newblock \emph{\bibinfo{journal}{Nature Communications}} \textbf{\bibinfo{volume}{14}}, \bibinfo{pages}{3499} (\bibinfo{year}{2023}).

\bibitem{nagarajan_large-scale_2005}
\bibinfo{author}{Nagarajan, R.} \emph{et~al.}
\newblock \bibinfo{title}{Large-scale photonic integrated circuits}.
\newblock \emph{\bibinfo{journal}{IEEE Journal of Selected Topics in Quantum Electronics}} \textbf{\bibinfo{volume}{11}}, \bibinfo{pages}{50--65} (\bibinfo{year}{2005}).

\bibitem{stojanovic_monolithic_2018}
\bibinfo{author}{Stojanović, V.} \emph{et~al.}
\newblock \bibinfo{title}{Monolithic silicon-photonic platforms in state-of-the-art {CMOS} {SOI} processes [{Invited}]}.
\newblock \emph{\bibinfo{journal}{Optics Express}} \textbf{\bibinfo{volume}{26}}, \bibinfo{pages}{13106--13121} (\bibinfo{year}{2018}).

\bibitem{sacher_multilayer_2015}
\bibinfo{author}{Sacher, W.~D.}, \bibinfo{author}{Huang, Y.}, \bibinfo{author}{Lo, G.-Q.} \& \bibinfo{author}{Poon, J. K.~S.}
\newblock \bibinfo{title}{Multilayer {Silicon} {Nitride}-on-{Silicon} {Integrated} {Photonic} {Platforms} and {Devices}}.
\newblock \emph{\bibinfo{journal}{Journal of Lightwave Technology}} \textbf{\bibinfo{volume}{33}}, \bibinfo{pages}{901--910} (\bibinfo{year}{2015}).

\bibitem{bose_anneal-free_2024}
\bibinfo{author}{Bose, D.} \emph{et~al.}
\newblock \bibinfo{title}{Anneal-free ultra-low loss silicon nitride integrated photonics}.
\newblock \emph{\bibinfo{journal}{Light: Science \& Applications}} \textbf{\bibinfo{volume}{13}}, \bibinfo{pages}{156} (\bibinfo{year}{2024}).

\bibitem{lu2014selective}
\bibinfo{author}{Lu, X.}, \bibinfo{author}{Rogers, S.}, \bibinfo{author}{Jiang, W.~C.} \& \bibinfo{author}{Lin, Q.}
\newblock \bibinfo{title}{Selective engineering of cavity resonance for frequency matching in optical parametric processes}.
\newblock \emph{\bibinfo{journal}{Applied Physics Letters}} \textbf{\bibinfo{volume}{105}}, \bibinfo{pages}{151104} (\bibinfo{year}{2014}).

\bibitem{yu2021spontaneous}
\bibinfo{author}{Yu, S.-P.} \emph{et~al.}
\newblock \bibinfo{title}{Spontaneous pulse formation in edgeless photonic crystal resonators}.
\newblock \emph{\bibinfo{journal}{Nature Photonics}} \textbf{\bibinfo{volume}{15}}, \bibinfo{pages}{461--467} (\bibinfo{year}{2021}).

\bibitem{black_optical-parametric_2022}
\bibinfo{author}{Black, J.~A.} \emph{et~al.}
\newblock \bibinfo{title}{Optical-parametric oscillation in photonic-crystal ring resonators}.
\newblock \emph{\bibinfo{journal}{Optica}} \textbf{\bibinfo{volume}{9}}, \bibinfo{pages}{1183} (\bibinfo{year}{2022}).

\bibitem{zang_foundry_2024}
\bibinfo{author}{Zang, J.}, \bibinfo{author}{Liu, H.}, \bibinfo{author}{Briles, T.~C.} \& \bibinfo{author}{Papp, S.~B.}
\newblock \bibinfo{title}{Foundry manufacturing of octave-spanning microcombs}.
\newblock \emph{\bibinfo{journal}{Optics Letters}} \textbf{\bibinfo{volume}{49}}, \bibinfo{pages}{5143--5146} (\bibinfo{year}{2024}).

\bibitem{jin_bandgap-detuned_2025}
\bibinfo{author}{Jin, Y.} \emph{et~al.}
\newblock \bibinfo{title}{The bandgap-detuned excitation regime in photonic-crystal resonators}.
\newblock \emph{\bibinfo{journal}{Nature Communications}} \textbf{\bibinfo{volume}{16}}, \bibinfo{pages}{5077} (\bibinfo{year}{2025}).

\bibitem{black_group-velocity-dispersion_2021}
\bibinfo{author}{Black, J.~A.} \emph{et~al.}
\newblock \bibinfo{title}{Group-velocity-dispersion engineering of tantala integrated photonics}.
\newblock \emph{\bibinfo{journal}{Optics Letters}} \textbf{\bibinfo{volume}{46}}, \bibinfo{pages}{817} (\bibinfo{year}{2021}).

\bibitem{gayer_temperature_2008}
\bibinfo{author}{Gayer, O.}, \bibinfo{author}{Sacks, Z.}, \bibinfo{author}{Galun, E.} \& \bibinfo{author}{Arie, A.}
\newblock \bibinfo{title}{Temperature and wavelength dependent refractive index equations for {MgO}-doped congruent and stoichiometric {LiNbO3}}.
\newblock \emph{\bibinfo{journal}{Applied Physics B}} \textbf{\bibinfo{volume}{91}}, \bibinfo{pages}{343--348} (\bibinfo{year}{2008}).

\bibitem{carollo_amorphous_2025}
\bibinfo{author}{Carollo, A.~R.} \emph{et~al.}
\newblock \bibinfo{title}{Amorphous metal oxide mixtures for high-{Q} integrated nonlinear photonics}
\newblock \bibinfo{note}{ArXiv:2508.14887 [physics]}.

\bibitem{rao_actively-monitored_2019}
\bibinfo{author}{Rao, A.} \emph{et~al.}
\newblock \bibinfo{title}{Actively-monitored periodic-poling in thin-film lithium niobate photonic waveguides with ultrahigh nonlinear conversion efficiency of 4600\%{W}$^{\textrm{-1}}$cm$^{\textrm{-2}}$}.
\newblock \emph{\bibinfo{journal}{Optics Express}} \textbf{\bibinfo{volume}{27}}, \bibinfo{pages}{25920--25930} (\bibinfo{year}{2019}).

\bibitem{fazio_structure_2020}
\bibinfo{author}{Fazio, M.~A.} \emph{et~al.}
\newblock \bibinfo{title}{Structure and morphology of low mechanical loss {TiO}$_{\textrm{2}}$-doped {Ta}$_{\textrm{2}}${O}$_{\textrm{5}}$}.
\newblock \emph{\bibinfo{journal}{Optical Materials Express}} \textbf{\bibinfo{volume}{10}}, \bibinfo{pages}{1687--1703} (\bibinfo{year}{2020}).

\bibitem{yi_soliton_2015}
\bibinfo{author}{Yi, X.}, \bibinfo{author}{Yang, Q.-F.}, \bibinfo{author}{Yang, K.~Y.}, \bibinfo{author}{Suh, M.-G.} \& \bibinfo{author}{Vahala, K.}
\newblock \bibinfo{title}{Soliton frequency comb at microwave rates in a high-{Q} silica microresonator}.
\newblock \emph{\bibinfo{journal}{Optica}} \textbf{\bibinfo{volume}{2}}, \bibinfo{pages}{1078} (\bibinfo{year}{2015}).

\bibitem{lu_milliwatt-threshold_2019}
\bibinfo{author}{Lu, X.} \emph{et~al.}
\newblock \bibinfo{title}{Milliwatt-threshold visible–telecom optical parametric oscillation using silicon nanophotonics}.
\newblock \emph{\bibinfo{journal}{Optica}} \textbf{\bibinfo{volume}{6}}, \bibinfo{pages}{1535--1541} (\bibinfo{year}{2019}).

\bibitem{roberts_measurements_2022}
\bibinfo{author}{Roberts, S.}, \bibinfo{author}{Ji, X.}, \bibinfo{author}{Cardenas, J.}, \bibinfo{author}{Corato-Zanarella, M.} \& \bibinfo{author}{Lipson, M.}
\newblock \bibinfo{title}{Measurements and {Modeling} of {Atomic}-{Scale} {Sidewall} {Roughness} and {Losses} in {Integrated} {Photonic} {Devices}}.
\newblock \emph{\bibinfo{journal}{Advanced Optical Materials}} \textbf{\bibinfo{volume}{10}}, \bibinfo{pages}{2102073} (\bibinfo{year}{2022}).

\bibitem{brodnik_nanopatterned_2025}
\bibinfo{author}{Brodnik, G.~M.}, \bibinfo{author}{Liu, H.}, \bibinfo{author}{Carlson, D.~R.}, \bibinfo{author}{Black, J.~A.} \& \bibinfo{author}{Papp, S.~B.}
\newblock \bibinfo{title}{Nanopatterned parametric oscillators for laser conversion beyond an octave}.
\newblock \emph{\bibinfo{journal}{Optica}} \textbf{\bibinfo{volume}{12}}, \bibinfo{pages}{337--342} (\bibinfo{year}{2025}).

\bibitem{domeneguetti_parametric_2021}
\bibinfo{author}{Domeneguetti, R.~R.} \emph{et~al.}
\newblock \bibinfo{title}{Parametric sideband generation in {CMOS}-compatible oscillators from visible to telecom wavelengths}.
\newblock \emph{\bibinfo{journal}{Optica}} \textbf{\bibinfo{volume}{8}}, \bibinfo{pages}{316} (\bibinfo{year}{2021}).

\bibitem{sayson_octave-spanning_2019}
\bibinfo{author}{Sayson, N. L.~B.} \emph{et~al.}
\newblock \bibinfo{title}{Octave-spanning tunable parametric oscillation in crystalline kerr microresonators}.
\newblock \emph{\bibinfo{journal}{Nature Photonics}} \textbf{\bibinfo{volume}{13}}, \bibinfo{pages}{701--706} (\bibinfo{year}{2019}).

\bibitem{zang_laser_2025}
\bibinfo{author}{Zang, J.} \emph{et~al.}
\newblock \bibinfo{title}{Laser power consumption of soliton formation in a bidirectional {Kerr} resonator}.
\newblock \emph{\bibinfo{journal}{Nature Photonics}} \textbf{\bibinfo{volume}{19}}, \bibinfo{pages}{510--517} (\bibinfo{year}{2025}).

\bibitem{chen_adapted_2024}
\bibinfo{author}{Chen, P.-K.} \emph{et~al.}
\newblock \bibinfo{title}{Adapted poling to break the nonlinear efficiency limit in nanophotonic lithium niobate waveguides}.
\newblock \emph{\bibinfo{journal}{Nature Nanotechnology}} \textbf{\bibinfo{volume}{19}}, \bibinfo{pages}{44--50} (\bibinfo{year}{2024}).

\bibitem{xin_wavelength-accurate_2025}
\bibinfo{author}{Xin, C.~J.} \emph{et~al.}
\newblock \bibinfo{title}{Wavelength-accurate and wafer-scale process for nonlinear frequency mixers in thin-film lithium niobate}.
\newblock \emph{\bibinfo{journal}{Communications Physics}} \textbf{\bibinfo{volume}{8}}, \bibinfo{pages}{136} (\bibinfo{year}{2025}).

\bibitem{yu_continuum_2022}
\bibinfo{author}{Yu, S.-P.}, \bibinfo{author}{Lucas, E.}, \bibinfo{author}{Zang, J.} \& \bibinfo{author}{Papp, S.~B.}
\newblock \bibinfo{title}{A continuum of bright and dark-pulse states in a photonic-crystal resonator}.
\newblock \emph{\bibinfo{journal}{Nature Communications}} \textbf{\bibinfo{volume}{13}}, \bibinfo{pages}{3134} (\bibinfo{year}{2022}).

\bibitem{fujii2020dispersion}
\bibinfo{author}{Fujii, S.} \& \bibinfo{author}{Tanabe, T.}
\newblock \bibinfo{title}{Dispersion engineering and measurement of whispering gallery mode microresonator for kerr frequency comb generation}.
\newblock \emph{\bibinfo{journal}{Nanophotonics}} \textbf{\bibinfo{volume}{9}}, \bibinfo{pages}{1087--1104} (\bibinfo{year}{2020}).

\bibitem{malitson_interspecimen_1965}
\bibinfo{author}{Malitson, I.~H.}
\newblock \bibinfo{title}{Interspecimen {Comparison} of the {Refractive} {Index} of {Fused} {Silica}}.
\newblock \emph{\bibinfo{journal}{JOSA}} \textbf{\bibinfo{volume}{55}}, \bibinfo{pages}{1205--1209} (\bibinfo{year}{1965}).

\bibitem{boyd_nonlinear_optics}
\bibinfo{author}{Boyd, R.~W.}, \emph{\bibinfo{title}{Nonlinear Optics}}.
\newblock  (\bibinfo{publisher}{\bibinfo{edition}{4th} edn}, \bibinfo{year}{2020}).

\end{thebibliography}

\clearpage

\end{document}